# The Impact of Magnons, Defects, and Rapid Energy Migration on the Optical Properties of the 2D Magnet $CrPS_4$

Jacob T. Baillie,[1] Eden Tzanetopoulos,[1] Rachel T. Smith,[1]
Remi Beaulac,[2] and Daniel R. Gamelin[1,*]
[1]*Department of Chemistry, University of Washington, Seattle, WA 98195*
[2]*Department of Chemistry, Swarthmore College, Swarthmore, PA 19081*
*Contact author: gamelin@uw.edu

**Abstract.** Strong coupling between optical and magnetic excitations could enable contactless, spatially resolved, or ultrafast interrogation and control of magnetism in two-dimensional (2D) materials and devices. The layered 2D A-type antiferromagnet $CrPS_4$ stands out among van der Waals (vdW) magnets for its rich optical fine structure, but its spectroscopy is not yet understood and has so far been interpreted without consideration of magnetic exchange. Here, we show that this fine structure comes primarily from exchange-mediated coupling between on-site optical "spin-flip" transitions of $Cr^{3+}$ and low-energy spin transitions involving the surrounding lattice. Well-resolved magnon sidebands to optical $^4A_2 \leftrightarrow {}^2E$ transitions are observed in photoluminescence (PL) and PL excitation spectra, as well as a pronounced PL sideband due to short-range exchange splitting. Energy migration is probed using $Yb^{3+}$ dopants as traps, revealing sub-picosecond inter-site excitation hopping. Formation of dispersive Frenkel excitons of coupled on-site *d-d* transitions due to inter-site exchange is discussed. In addition to impacting how optical fine structure is interpreted in this and potentially other vdW magnets, these findings may have ramifications for future applications of layered 2D magnets by revealing new opportunities to drive mode-specific spin-wave excitations using light.

## I. Introduction

The emergence of two-dimensional (2D) van der Waals (vdW) magnets has stimulated advances in the fundamental physics of strongly correlated systems as well as the development of nascent nanoscale spintronic technologies.[1,2] Magnetic vdW materials have proven to be effective platforms for exploring purely magnetic phenomena, including layer-dependent magnetism,[3,4] the generation or transport of spin waves,[5-9] and spin-dependent tunneling.[10-12] Intrinsic coupling between optical and magnetic properties in such materials has further introduced avenues for contactless, spatially resolved, or ultrafast interrogation of magnetism[3,13-20] and, in some cases, optical control of spin orientation.[21-23] In three-dimensional magnetic materials, such coupling has already been used for spin-photonic transduction in optical isolators, magnetic sensors, and memory devices.[24-26] 2D magnets can offer analogous functionality with thicknesses down to atomic length scales in a form that is readily interfaced with incommensurate materials through hetero-stacking. In some 2D magnets, such as the workhorse $CrX_3$ (X = Cl, Br, I) compounds, optical transitions are broadened by strong electron-phonon coupling, obscuring the interplay between the optical and magnetic properties. In other materials, such as $CrPS_4$ and CrSBr, narrow optical transitions are observed with complex fine structure that is not fully understood. A deeper fundamental understanding of the optical properties and electronic structures of such materials will enable their full potential to be tapped for future applications.

Here, we present a detailed description of the optical properties of the A-type layered antiferromagnet, $CrPS_4$. $CrPS_4$ is an emerging 2D material whose unique magnetic and optical properties have only recently begun to be understood and exploited.[27-36] Figure 1a depicts the stacked lattice structure of $CrPS_4$ and the structure of one individual layer. The individual layer is highly anisotropic, with linear chains of edge-sharing $Cr^{3+}$ octahedra running along the crystallographic *b* axis. Parallel chains are separated by the S–P–S linkages of bridging $[PS_4]^{3-}$ moieties. Magnetic susceptibility[31,32] and neutron diffraction[30] experiments on bulk $CrPS_4$ have established A-type antiferromagnetic (A-AFM) ordering below a Néel temperature ($T_N$) of ~36 K, in which each monolayer orders ferromagnetically (FM) with all $Cr^{3+}$ moments oriented nearly perpendicular to the 2D plane, and these monolayers stack with opposite spin orientations to form the extended A-AFM lattice. Its structural anisotropy lends $CrPS_4$ a quasi-one-dimensional (1D) electronic structure that is manifested in disproportionately strong *b*-axis magnetic exchange coupling,[30] anisotropic magnon[8,9] and charge[37] transport, and primarily *b*-polarized absorption,[33,38] reflectance,[28] and PL[29] spectra. Interplay between the magnetic and optical properties of $CrPS_4$ has been found in magneto-PL experiments[29,36] and sensitivity of the PL intensity to temperature near $T_N$.[29,34]

Despite these promising results, the optical properties and electronic structure of $CrPS_4$ remain poorly defined, particularly pertaining to the luminescent excited state. Several[28,29,34-36,39-44] publications have interpreted the absorption, PL, reflectance, and photoconductivity spectra in terms of localized ligand-field electronic transitions from individual $Cr^{3+}$ ions (illustrated in Figure 1b). The low-temperature PL of $CrPS_4$ occurs at ~10970 $cm^{-1}$ (~1.3600 eV) and has been assigned to $^2E \rightarrow {}^4A_2$ ligand-field emission,[34,35] but the decay time of this PL is reportedly < 1 ns,[35] whereas $^2E$ emission typically decays on micro- to millisecond timescales due to the transition's spin-forbiddenness. A pronounced vibronic progression in the PL spectrum has been described,[29,34-36] and additional unusual fine structure has been attributed[29,34,43] in part to a Fano-type resonant coupling of the $^2E \rightarrow {}^4A_2$ transition with a separate atom-like oscillator, proposed to be excess atomic phosphorous within the lattice. Critically, the impact of magnetic exchange on the $CrPS_4$ PL spectrum has not been addressed. Additionally, to date, no absorption or PL excitation (PLE) spectra have been reported for the $^4A_2 \rightarrow {}^2E$ energy region that could test current interpretations.

In this study, we analyze the PL spectra of $CrPS_4$, complemented by absorption, high-resolution PLE, PL quantum yield (PLQY), and time-resolved PL data, and we present a revised interpretation of the fine structure of these spectra by explicitly accounting for magnetic exchange. We identify the electronic origins of the $^4A_2 \leftrightarrow {}^2E$ transitions in PL and PLE measurements at $E(^2E) \approx 11069$ cm$^{-1}$ (1.3724 eV) based on mirrored fine structure. From PL quantum-yield measurements, we are able to attribute the unusually fast $^2E$ decay to rapid nonradiative trapping, and using time-resolved PL, we identify emission from a native "defect" $Cr^{3+}$ site with a microsecond lifetime. Using $Yb^{3+}$ dopants as "designer" defects for capturing excitation energy from $CrPS_4$, time-resolved and variable-temperature PL measurements allow the conclusion of sub-picosecond site-to-site energy-hopping times, indicating that this excited state is a weakly dispersive Frenkel exciton. We identify several resolved magnon sidebands in the PL and PLE fine structure at low temperature, showing that exciton-magnon coupling is an important source of $^2E \leftrightarrow {}^4A_2$ PL and PLE intensity. Finally, we observe an additional relatively intense PL feature that is redshifted from the first $^2E \rightarrow {}^4A_2$ electronic origin and that can be assigned to a spin-conserving electronic transition terminating at an excited spin level of the exchange-split ground state. This transition serves as the origin of the most prominent phonon replicas in the PL spectra of $CrPS_4$. This work helps to establish $CrPS_4$ as a model system for the fundamental understanding of spin effects in the optical spectra of magnetic van der Waals compounds. Furthermore, the identification of well-resolved magnetic sidebands in $CrPS_4$ absorption and PL spectra reveals new opportunities to explore and harness coupling between optical excitations and spin waves in layered magnetic materials.

## II. Results and Analysis

**General Characterization.** $CrPS_4$ crystals were grown *via* chemical vapor transport. X-ray diffraction (XRD) and Raman spectroscopy data collected on single-crystal flakes align well with reference data (Figure S1). In the XRD pattern, the (00*l*) peaks are relatively more intense than in the reference pattern, reflecting orientation of the flake with its *c*-axis normal to the substrate. Field-cooled (FC) and zero-field-cooled (ZFC) magnetic susceptibility ($\chi$) measurements (Figure S2) confirm the A-type antiferromagnetism (A-AFM) reported previously.[31] Further details are provided in the Methods section and as Supplementary Material.

**Spectroscopic Overview.** Figure 1c shows unpolarized room-temperature and 18 K absorption spectra collected through a thin (52 ± 7 μm) $CrPS_4$ single crystal, with light propagating along the crystallographic *c* axis. Molar extinction coefficients ($\varepsilon$) were determined using the Beer-Lambert law based on the flake thickness, the $Cr^{3+}$ density in the lattice, and the measured optical density (Figure S3). The broad absorption band centered at ~14000 $cm^{-1}$ (~1.7360 eV, at 18 K) is assigned to the spin-allowed $^4A_2 \rightarrow {}^4T_2$ ligand-field transition of the pseudo-octahedral $Cr^{3+}$. The peak extinction coefficient of this band is 16 ± 2 $M^{-1}cm^{-1}$, which is reasonable for octahedral $Cr^{3+}$.[45] A similar extinction coefficient was measured for an ensemble of $CrPS_4$ microcrystals prepared by sonication (Figure S3). This band narrows when the crystal is cooled to 18 K, attributed to the loss of vibronic hot bands. The absorption maximum shifts to higher energy at lower *T*, consistent with the known Cr–S bond-length contraction in $CrPS_4$.[30,46] A series of more intense bands is observed at higher energies in the sonicated $CrPS_4$ microcrystals (Figure S3), with strong absorption beginning above ~18000 $cm^{-1}$ (~2.2317 eV). The first peak maximum in this region occurs at ~19500 $cm^{-1}$ (~2.4177 eV) and is assigned to a ligand-to-metal charge-transfer (LMCT) transition.[39]

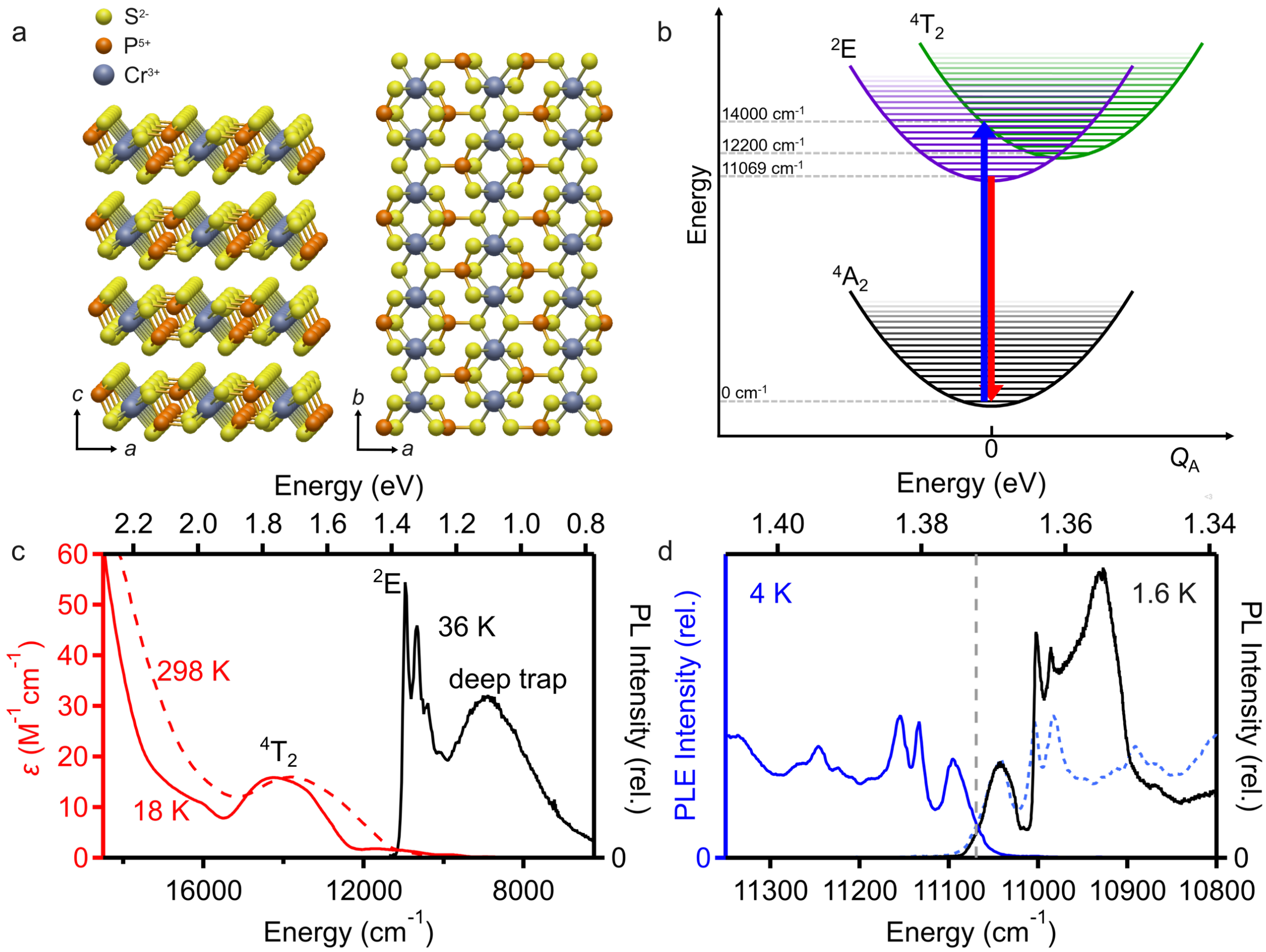

**Figure 1. (a)** Images of the structure of $CrPS_4$ based on the data in ref 46. **(b)** Configurational coordinate diagram (along the totally symmetric nuclear coordinate) representing the localized *d*-*d* transitions of an octahedral $d^3$ $Cr^{3+}$ ion in $CrPS_4$ at 4 K. The blue arrow represents one of the vibronic transitions of the $^4A_2 \rightarrow {}^4T_2$ absorption and the red arrow represents the electronic origin of the $^2E \rightarrow {}^4A_2$ luminescence. **(c)** Absorption spectra of a 52 ± 7 μm thick $CrPS_4$ single crystal (~70k layers, Figure S3) at 298 K (dashed red) and 18 K (solid red), collected using unpolarized light propagating along the crystallographic *c* axis. Extinction coefficients are estimated to be accurate to within ±14%. 36 K PL spectrum of a $CrPS_4$ single crystal (black) collected using pulsed 21277 $cm^{-1}$ photoexcitation (470 nm, unpolarized, 2.5 MHz, 3 ns pulse duration, 6.3 nJ pulse energy). A 1 nm (~10 $cm^{-1}$) spectral bandwidth was used for detection. **(d)** 1.6 K PL spectrum of a $CrPS_4$ single crystal (black) and 4 K PLE spectrum of a 0.2% $Yb^{3+}$:$CrPS_4$ single crystal (blue). The PL spectrum was collected using continuous-wave (CW) 15803 $cm^{-1}$ photoexcitation (633 nm, unpolarized, 5 mW/$cm^2$). A 0.05 nm (~0.5 $cm^{-1}$) spectral bandwidth was used for detection. The PLE spectrum was collected by photoexciting with a tunable, CW Ti:sapphire laser (unpolarized, 5–50 mW/$cm^2$, 0.1 nm linewidth) while monitoring PL from $Yb^{3+}$ dopants at 9302 $cm^{-1}$ (spectral bandwidth = ±22 $cm^{-1}$, see Figure S9). For reference, the PLE spectrum is also shown flipped (dotted light blue) across the electronic origin (dashed gray line at 11069 $cm^{-1}$ (1.3724 eV)).

Figure 1c also shows the 36 K PL spectrum of a $CrPS_4$ single crystal. A series of evenly spaced peaks is observed near 10500 $cm^{-1}$ (~1.30 eV) that has been attributed to vibronic coupling in the localized $^2E$ excited state.[29,34-36] The broad feature near 9000 $cm^{-1}$ (~1.10 eV) has been assigned to deep-trap PL.[23,35] The $^2E$ PL intensity is nearly linearly dependent on excitation power ($I \propto P^{1.08}$, Figure S4), whereas this deep-trap PL shows $I \propto P^{0.76}$ (Figure S4), indicating saturation. The relatively low excitation power densities used in this work (< 2 $W/cm^2$) compared with the previous report of this feature (~3 $kW/cm^2$)[35] may explain why this deep-trap PL appears relatively more intense here. Overall, the 36 K PL spectrum is consistent with literature reports. The results and analysis presented below focus on understanding the $CrPS_4$ PL and absorption (or PLE) spectra in the region of the first electronic excitations.

**Low-Temperature Absorption (or PLE) and PL.** Figure 1d shows the PL spectrum of a $CrPS_4$ single crystal measured at 1.6 K together with a 4 K laser PLE spectrum, expanded around the highest-energy $^2E$ PL peak of Figure 1c. The PL spectrum shows extensive fine structure, as discussed previously,[29,34,36,43,44] with its first three peak maxima at 11042, 11002, and 10986 $cm^{-1}$ (1.3690, 1.3641, 1.3621 eV). This fine structure shows negligible dependence on excitation position within the crystal (Figure S5), excitation energy (Figures S6 and S7), or excitation power (Figure S6), and it shows very little flake-to-flake variability (Figure S8). The same PL fine structure is observed in bulk samples (*e.g.*, Figure 1d),[29,43] thin exfoliated (~100 nm) samples,[34,44] and microcrystals (~10 nm thick, Figures S3 and S17, inhomogeneously broadened), indicating this fine structure is not strongly affected by normal sample-dependent parameters such as thickness, disorder, strain, *etc*.

The $CrPS_4$ PLE spectrum was collected at 4 K using a 0.2% $Yb^{3+}$-doped $CrPS_4$ ($Yb^{3+}$:$CrPS_4$) single crystal (see Figure S9 for PL). Although there is no significant difference in $CrPS_4$ PLE fine structure near the PL origin between $Yb^{3+}$:$CrPS_4$ and $CrPS_4$ samples (Figure S10), $Yb^{3+}$ doping provides much greater PL quantum yields[23] and consequently greater signal-to-noise in the PLE measurements. As illustrated in Figure 1d, the PLE spectrum mostly resembles the PL spectrum mirrored across a midpoint energy of 11069 $cm^{-1}$ (1.3724 eV, dashed gray line in Figure 1d, see Figure S30 and discussion below), with its first three peak maxima at 11096, 11134, and 11155 $cm^{-1}$ (1.3757, 1.3804, 1.3831 eV). These data indicate that the fine structure in both excitation and emission is built upon a common pure-electronic origin at ca. 11069 $cm^{-1}$ (1.3724 eV). Absorption at this energy is very weak, with $\varepsilon$ < ~0.5 $M^{-1}cm^{-1}$ (Figure S11). This small extinction coefficient is characteristic of spin-forbidden ligand-field transitions of $Cr^{3+}$ and supports the assignment of this absorption and PL to the $^4A_2 \leftrightarrow {}^2E$ electronic transition of $Cr^{3+}$. The more intense, broader absorption band centered at ~14000 $cm^{-1}$ (~1.7358 eV, Figure 1c) is assigned as the first spin-allowed ligand-field transition, $^4A_2 \rightarrow {}^4T_2$, broadened slightly by the presence of two $Cr^{3+}$ lattice sites and the fact that each has low symmetry. This band is also observed in extended PLE spectra (see Supplementary Material, Figures S10 and S11). Relevant energies are included in Figure 1b.

The observation of $^2E$ absorption before the onset of $^4T_2$ absorption demonstrates that $Cr^{3+}$ in $CrPS_4$ resides in a strong ligand-field environment at 4 K. From ligand-field theory for octahedral $Cr^{3+}$, $E(^4T_2) = 10Dq$, such that $Dq = 1400$ $cm^{-1}$ in $CrPS_4$. Diagonalization of the 4×4 matrix of $^2E$ energies gives as the lowest root $E(^2E)/B \approx 3.05\ C/B + 7.09 - 1.80\ B/Dq$,[47,48] where $B$ and $C$ are the Racah parameters describing electron-electron repulsion. No other ligand-field excited states could be identified below the charge-transfer onset, so $B$ and $C$ cannot both be determined experimentally. Using the experimental energy $E(^2E) = 11069$ $cm^{-1}$ and a typical ratio of $C/B = 4.5$ for $Cr^{3+}$ compounds yields $B = 551$ $cm^{-1}$ and hence $Dq/B \sim 2.54$. This value is greater than previously estimated ($Dq/B \sim 2.08$),[34] and it demonstrates that $Cr^{3+}$ is further from the excited-state spin-crossover point at 4 K than previously thought. Perhaps more importantly, the large reduction of $B$ from its free-ion value ($B_0 = 933$ $cm^{-1}$)[49] has not yet been commented upon. This reduction results from $Cr^{3+}$ 3$d$ covalency, which diminishes electron-electron repulsion through the nephelauxetic effect.[45,50] Large ranges in the value of $C/B$ for $Cr^{3+}$ ions have been quoted in recent literature,[51] and consequently the usual nephelauxetic parameter $\beta = B/B_0$ appears poorly correlated with $E(^2E)$ in many $Cr^{3+}$ compounds.[52] To minimize uncertainty from assumption of a specific $C/B$ ratio, we may instead express the reduction of electron-electron repulsion in terms of the "new" nephelauxetic parameter, $\beta_1$, described in ref 52 and defined in eq 1a (using free-ion $B_0$ = 933 $cm^{-1}$ and $C_0 = 3732$ $cm^{-1}$ from ref 49).

$$\beta_1 = \sqrt{\left(\frac{B}{B_0}\right)^2 + \left(\frac{C}{C_0}\right)^2} \qquad (1a)$$

To normalize relative to the free ion, we redefine this parameter as in eq 1b. Note that if $C/B = C_0/B_0$ then $\kappa = \beta$.

$$\kappa = \frac{1}{\sqrt{2}}\sqrt{\left(\frac{B}{B_0}\right)^2 + \left(\frac{C}{C_0}\right)^2} \quad (1b)$$

For $CrPS_4$ at 4 K, $\kappa = 0.62$, which is essentially independent of the assumed *C/B* ratio. This value of $\kappa$ is low even for sulfide-coordinated $Cr^{3+}$. For example, $\kappa = 0.67$ for $Cr^{3+}$ doped into the thiospinel, $ZnAl_2S_4$.[53] Consequently, the $^2E$ energy is lower in $CrPS_4$ than in $Cr^{3+}$-doped $ZnAl_2S_4$ (11069 *vs* 12970 $cm^{-1}$, or 1.3724 *vs* 1.6081 eV). The high $Cr^{3+}$ *d*-orbital covalency in $CrPS_4$ indicated by this result is attributable to the high polarizability of the $[PS_4]^{3-}$ anion and the low coordination number of the sulfides that occupy the out-of-plane $Cr^{3+}$ coordination sites; each sulfide in $ZnAl_2S_4$ is 4-coordinate, whereas in $CrPS_4$ the four bridging sulfides are 3-coordinate and the two out-of-plane sulfides are only 2-coordinate. With the above ligand-field parameters ($Dq$ = 1400 $cm^{-1}$, $B$ = 551 $cm^{-1}$, $C/B$ = 4.5), the $^4A_2 \rightarrow {}^4T_1(F)$ transition in $CrPS_4$ is expected to occur at ~19500 $cm^{-1}$, where it would be obscured by charge-transfer absorption (Figure 1c).

**Variable-Temperature $^2$E Absorption (PLE) and PL.** Figure 2 replots the low-temperature PL spectrum of Figure 1d together with additional PL and PLE data collected at various temperatures up to 105 K. In the low-temperature PL spectra, evenly spaced PL and PLE peaks are indicated with gray or light blue lines. The ~300 $cm^{-1}$ (~37.2 meV) spacing between these replicas can be attributed to the $A_6$ phonon mode (Figure S1b), whose Raman scattering is polarized in the *ab* plane.[54,55] The coupling is not strong, with the Huang-Rhys parameter estimated from the ratio of second to first peak intensities to be $S \sim 0.5$, depending on the choice of baseline. As the temperature increases, the PL and PLE peaks broaden and the relative peak intensities change dramatically, but the peak energies barely change (Figure S15). The 1.6 K PL spectrum is dominated by fine-structure features close to the electronic origin. This intensity decreases rapidly between 1.6 and 4 K (note the ×1/6 scaling factor for the 1.6 K data in Figure 2), leading to the emergence of three peaks near 10750 $cm^{-1}$ (1.3328 eV), well below the PL origin. Upon close inspection, these peaks are also evident in the 1.6 K PL spectrum. These three peaks show the same splitting and intensity pattern as the higher-energy peaks centered around 10960 $cm^{-1}$ (1.3589 eV)). In fact, each of these peaks occurs exactly 208 ± 1 $cm^{-1}$ (25.8 ± 0.1 meV) below the corresponding member of the first set (Figure S12). Moreover, 300 $cm^{-1}$ replicas associated with these new peaks are also identifiable (blue lines in Figure 2). Due to this nearly identical fine structure, we interpret this 10750 $cm^{-1}$ (1.3328 eV) PL as $^2E$ emission from a "defect" $Cr^{3+}$ site, an interpretation borne out by time-resolved PL measurements discussed below. The similar fine structure is expected because the $^2E \rightarrow {}^4A_2$ emission of such defect $Cr^{3+}$ sites can couple similarly to phonons and magnons (see below), as observed in, *e.g.*, $YCrO_3$.[56]

With this assignment, the strong temperature dependence of the PL intensities in Figure 2 can be understood: it stems from the different temperature dependencies of lattice and defect $Cr^{3+}$ luminescence, which in turn reflect the different nonradiative processes affecting each center. Specifically, the substantial decrease in relative intensity of the lattice $Cr^{3+}$ PL (10950 $cm^{-1}$ (1.3576 eV) and associated progression) upon warming from 1.6 to 4 K is attributed to thermally activated energy migration to killer traps. The defect $Cr^{3+}$ PL shows a different temperature dependence because of its trapping energy and the relatively low defect $Cr^{3+}$ concentration. By 36 K, the PL and PLE fine structure is no longer resolved; only the 300 $cm^{-1}$ (37.2 meV) progression of the lattice $Cr^{3+}$ is observed, and these features continue broadening with increasing temperature. Given the strong temperature dependence of this PL spectrum at low temperatures, the many inconsistencies among previously reported[23,29,34-36,43,44] low-temperature $CrPS_4$ PL spectra are likely primarily attributable to sample-to-sample variations in *nonradiative* processes (*e.g.*, from different trap and defect $Cr^{3+}$ densities), compounded by variations in laboratory sample temperature.

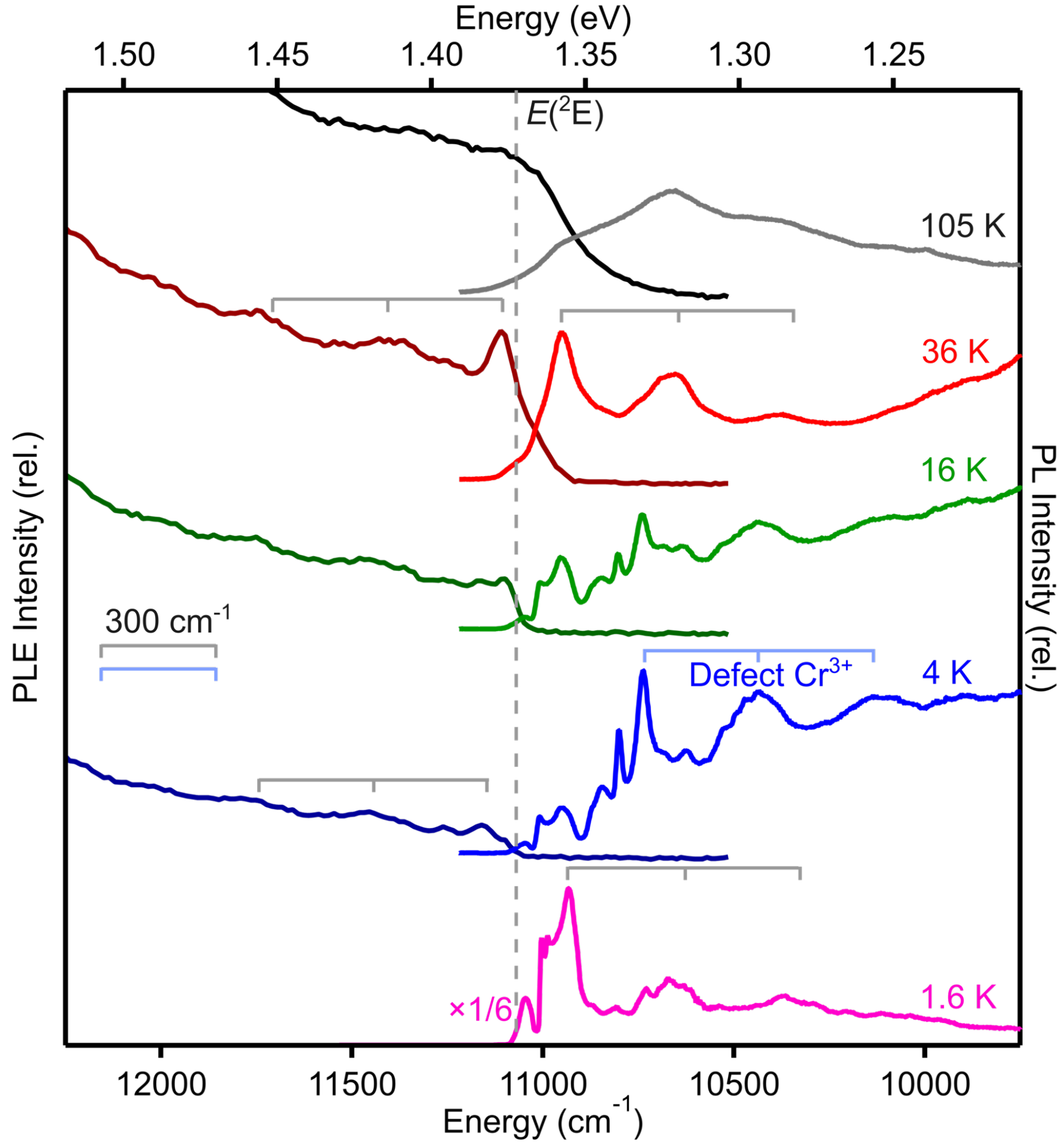

**Figure 2.** Variable-temperature PL spectra of $CrPS_4$ single crystals (lighter colors) and PLE spectra of a 0.2% $Yb^{3+}$:$CrPS_4$ single crystal (darker colors). PL and PLE spectra for $T$ = 4, 16, 36, 105 K are from the same single crystal measured in a closed-cycle cryostat. A different crystal was used for the $T$ = 1.6 K measurements in a helium immersion cryostat. Spectra were collected using CW 15803 $cm^{-1}$ photoexcitation (633 nm, unpolarized, 5 $mW/cm^2$). A 0.05 nm (~0.5 $cm^{-1}$) spectral bandwidth was used for detection. The dashed gray line indicates $E(^2E)$ = 11069 $cm^{-1}$ (1.3724 eV) in the low-temperature limit. Bars are drawn for the first two replicas of each component at select temperatures as a guide to the eye. The spacing is 300 $cm^{-1}$ (37.2 meV). PLE spectra were collected by photoexciting with a CW halogen lamp (10-50 μ$W/cm^2$, 1 nm linewidth) while monitoring PL of $Yb^{3+}$ dopants at 9302 $cm^{-1}$ (spectral bandwidth ±22 $cm^{-1}$, see Figure S9). Note the scaling factor (×1/6) for the 1.6 K data, estimated from variable-temperature PL spectra of the same crystal (Figure S31): the lattice $Cr^{3+}$ PL intensity (gray bars) decreases rapidly upon warming from 1.6 to 4 K but the defect $Cr^{3+}$ PL intensity (blue bars) is relatively constant.

An important consideration when assessing variable-temperature PL data is the PLQY ($\Phi$), which quantifies the competition between radiative and nonradiative relaxation processes. The total PLQY was measured for several $CrPS_4$ flakes at room temperature using an integrating sphere, yielding $\Phi_{\mathrm{tot}}$ = 0.14 ± 0.04%. From this value and full spectral measurements performed between room temperature and 4 K (Figures S13-S16), we can determine that $\Phi_{\mathrm{tot}}$ = ~0.5% below 60 K. Thus, the PLQY of $CrPS_4$ is very low at all temperatures. Furthermore, by analysis of the spectral data to omit the deep-trap and defect $Cr^{3+}$ (near 10750 $cm^{-1}$ (~1.3328 eV), *vide infra*) PL intensities, the PLQY for just the intrinsic $^2E \rightarrow {}^4A_2$ component of the PL spectrum can also be determined, giving $\Phi_{^2\mathrm{E}}$ < 0.1% at all temperatures. These results indicate that $CrPS_4$ photoexcitation is overwhelmingly followed by nonradiative deactivation, rather than by PL. Changes in absolute PL intensities with parameters such as temperature therefore most likely reflect changes in the dynamics of nonradiative processes and require cautious interpretation.

**Time-Resolved PL.** More can be learned from studying the PL spectrum in the time domain. Figure 3 shows time-resolved PL data measured in the $^2E \rightarrow {}^4A_2$ region at different temperatures. At 4 K, two PL components are identified that have very different decay times. The short-time data are dominated by a "Fast" component with $\tau_{\mathrm{fast}}$ ~ 2 ns and with intensity originating at ~10960 $cm^{-1}$ (1.3589 eV). At longer times, a much slower component is observed, originating at ~10750 $cm^{-1}$ (~1.3328 eV). Separate measurements with a slower excitation repetition rate show that this "Slow" PL has a decay time of $\tau_{\mathrm{slow}}$ ~ 7 μs at 4 K. The 2.5 nm (~25 $cm^{-1}$) spectral bandwidth used for detection in this experiment causes instrument-limited linewidths, but the features are easily correlated with those in the CW spectra measured at higher resolution (Figures 1 and 2). Based on its relationship to the PLE onset (Figure 2), the "Fast" PL corresponds to the majority species in $CrPS_4$, such that we can associate it with the $^2E$ luminescence from the lattice $Cr^{3+}$ ions ($\tau_{\mathrm{fast}} \equiv \tau_{^2\mathrm{E}}$). The absence of PLE intensity near the onset of the "Slow" PL (Figures 1d, 2) allows assignment of this "Slow" PL to a rarer "defect" $Cr^{3+}$ ($\tau_{\mathrm{slow}} \equiv \tau_{\mathrm{defectCr}}$). Defect PL in fully concentrated $Cr^{3+}$ lattices commonly stems from the occurrence of native defects that perturb the energies of proximal lattice $Cr^{3+}$ ions. Microcrystalline $CrPS_4$ powder prepared by sonication shows the same defect $Cr^{3+}$ PL with similar intensity relative to lattice $Cr^{3+}$ PL (Figure S17), suggesting that it is associated with native bulk defects rather than surface or edge defects.

The bottom panels of Figure 3 show time-gated PL spectra extracted by integrating the time-resolved data over the first 5 ns (0–5 ns, lattice $^2E$) and over a later 5 ns window (145–150 ns, defect $Cr^{3+}$). As noted above, the defect $Cr^{3+}$ spectrum is very similar to the lattice $^2E$ spectrum, just shifted to lower energy by 208 $cm^{-1}$ (25.8 meV) and having distinct temperature and time dependencies. Overall, these results demonstrate that the $CrPS_4$ PL spectrum derives some of its complexity from the presence of $^2E$ emission from two $Cr^{3+}$ species: regular lattice sites and defect sites.

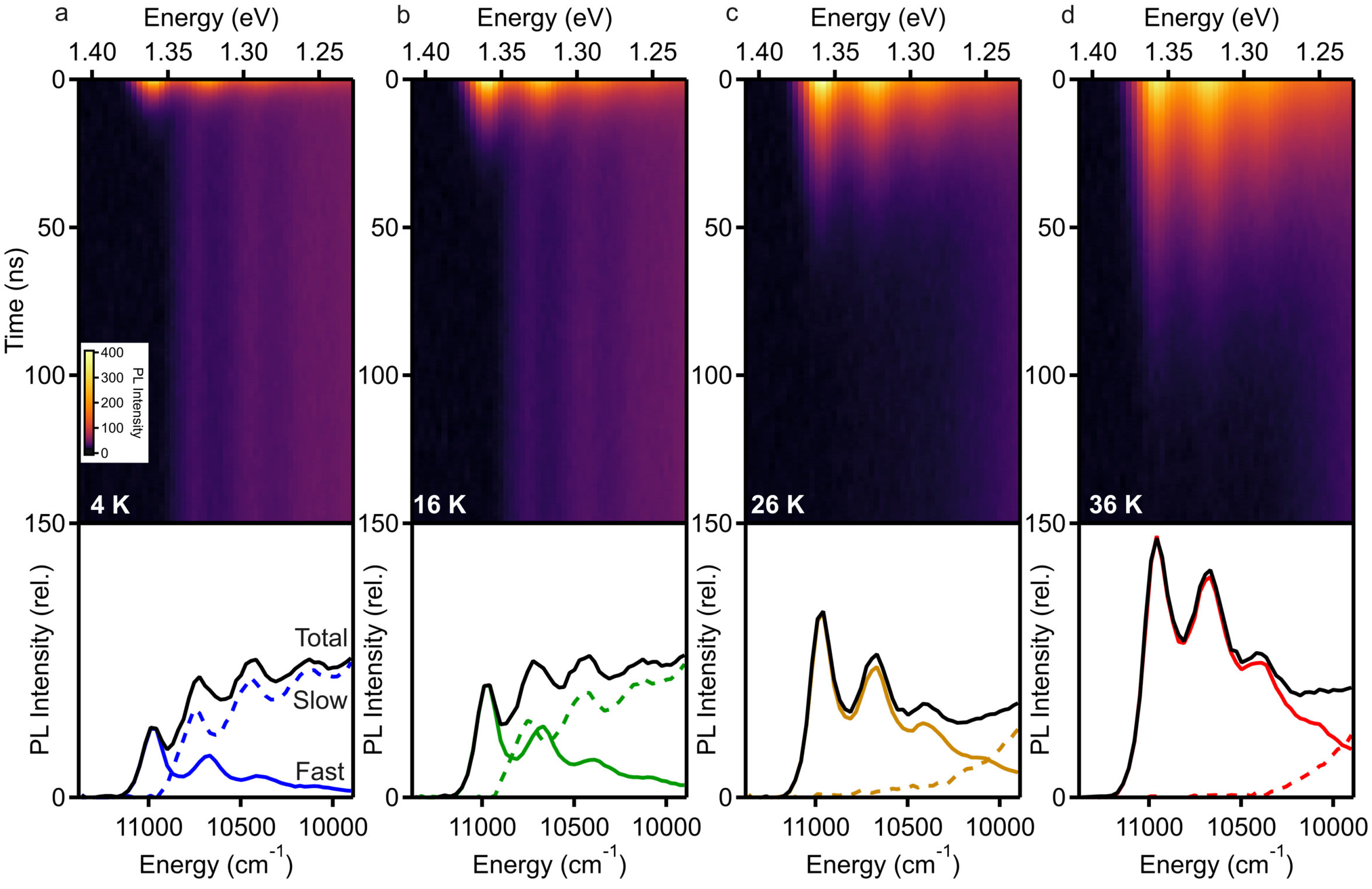

**Figure 3. (a)** Top: 4 K false-color plot of the PL decay *vs* energy measured across the region of the $^2E \rightarrow {}^4A_2$ transition with a 2.5 nm (~25 cm$^{-1}$) spectral bandwidth. Bottom: PL spectra obtained by integrating across all time ($I_{total}$, "Total", black), over the first 5 ns ($I_{0\text{-}5\,ns}$ - $I_{145\text{-}150\,ns}$, intrinsic lattice $^2E$, solid), and over a later 5 ns window ($I_{145\text{-}150\,ns}$, defect $Cr^{3+}$, dashed). The lattice $^2E$ and defect $Cr^{3+}$ spectra are scaled so their sum equals $I_{total}$. **(b-d)** Same as panel (a) but measured at 16, 26, and 36 K, respectively. All data are for a $CrPS_4$ single crystal and collected using pulsed 21277 cm$^{-1}$ photoexcitation (470 nm, 2.5 MHz, 3 ns pulse duration, 6.3 nJ pulse energy).

Figure 4 summarizes the decay dynamics of the intrinsic lattice $^2E$ and defect $Cr^{3+}$ PL components. The lattice $^2E$ PL decays monoexponentially with $\tau_{^2E}$ = 1.7 ns at 4 K, slowing to $\tau_{^2E}$ = 6.0 ns at 24 K (Figure 4a). The full temperature dependence of $\tau_{^2E}$ up to 60 K (Figure 4b) shows a gradual increase with increasing temperature over this range. The opposite temperature dependence is observed in Figure 4c for the slow defect $Cr^{3+}$ PL, whose decay accelerates from $\tau_{defectCr}$ ~ 7 μs at 4 K to 0.7 μs at 24 K. Figure 4d shows that $\tau_{defectCr}$ changes little at low temperatures up to ~16 K, at which point it drops rapidly with increasing temperature. Additional data detailing the distinct temperature dependencies of these two PL signals are provided in the Supplementary Materials (Figures S18-S20).

Using the PL quantum yields described above (Figure S16), it is possible to analyze these decay dynamics to estimate the radiative and nonradiative rate constants ($k_r$ and $k_{nr}$, respectively) for decay of the lattice $^2$E population at each temperature using eqs 2 and 3 (see Figure S21).

$$k_{\mathrm{r}} = \Phi_{^2\mathrm{E}} \cdot k_{^2\mathrm{E}} = \tau_{\mathrm{r}}^{-1} \tag{2}$$

$$k_{\mathrm{nr}} = k_{^2\mathrm{E}} - k_{\mathrm{r}} \tag{3}$$

This analysis suggests a radiative lifetime of $\tau_r$ = 7 μs at 4 K for the lattice $^2$E PL, which coincidentally is similar to the experimental value of $\tau_{defectCr}$ = 7 μs at 4 K. Although the very low PLQY may cause uncertainty in the precise radiative lifetime deduced here, slow radiative decay is already anticipated from the small extinction coefficients of the corresponding absorption transitions (Figure 1). Note that the low-temperature radiative decay is accelerated by spin effects that have no counterpart in the low-temperature absorption (see below), allowing faster radiative decay than anticipated from the absorption spectrum alone. Overall, this analysis shows that the short PL decay times widely observed near the absorption edge of $CrPS_4$ are primarily due to *nonradiative* processes, and the radiative lifetimes of these transitions are instead on the order of microseconds.

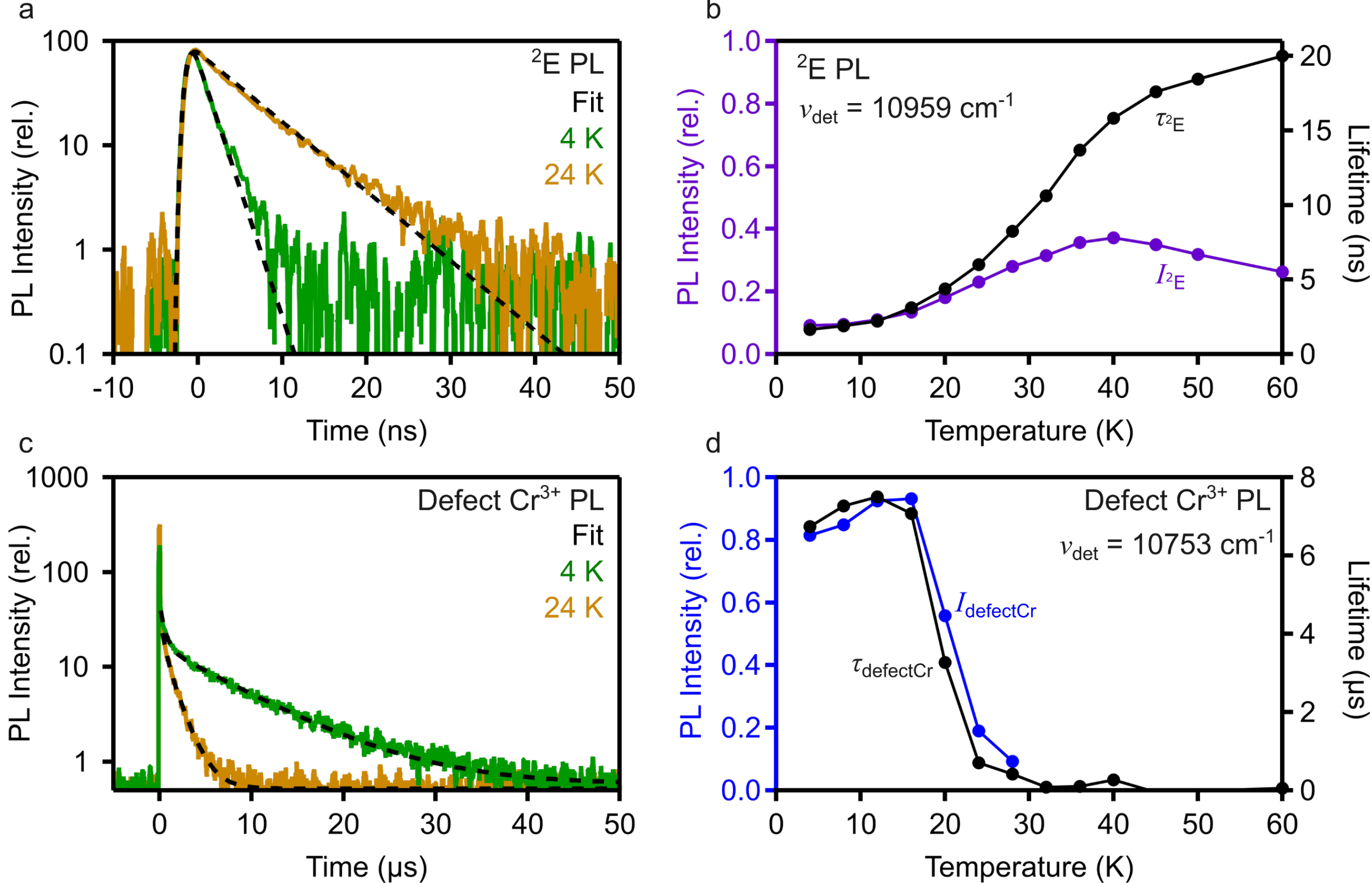

**Figure 4. (a)** 4 K (green) and 24 K (orange) PL decay curves of the intrinsic lattice $^2$E PL at a detection energy ($\nu_{\mathrm{det}}$) of 10959 cm$^{-1}$ (spectral bandwidth ±15 cm$^{-1}$) collected using pulsed 21277 cm$^{-1}$ photoexcitation (470 nm, unpolarized, 2.5 MHz, 3 ns pulse duration, 6.3 nJ pulse energy). $\tau_{^2\mathrm{E}}$ is determined by fitting to the convolution of the modeled monoexponential and the instrument response function (IRF) using nonlinear least-squares regression (dashed black, see Figure S18 for fitting details). **(b)** Temperature dependence of $I_{^2\mathrm{E}}$ from panel (a) (purple) and the corresponding PL decay time (black). **(c)** 4 K (green) and 24 K (orange) PL decay curves of the slow defect $Cr^{3+}$ PL at 10753 cm$^{-1}$ (spectral bandwidth ±14 cm$^{-1}$) collected by photoexciting with the pulsed 18797 cm$^{-1}$ (532 nm) output of a frequency-doubled Nd:YAG laser (unpolarized, 50 Hz, 30 ps pulse duration, 4 μJ pulse energy). The illustrated fits (dashed black) are biexponential decays starting from $t$ = 50 ns to avoid temporal overlap with the fast lattice $^2$E PL. $\tau_{\mathrm{defectCr}}$ is determined using eq S1. **(d)** Temperature dependence of $I_{\mathrm{defectCr}}$ from panel (c) (blue, integrated over 0.05–75 μs) and the corresponding defect $Cr^{3+}$ PL decay time (black).

With this consideration in mind, Figure 4d plots the temperature dependence of the defect $Cr^{3+}$ PL intensities ($I_{\mathrm{defectCr}}$, obtained by integrating decay data) together with the temperature dependence of the defect $Cr^{3+}$ PL decay times, for comparison with Figure 4b. Similar trends for $I_{^2\mathrm{E}}$ and $I_{\mathrm{defectCr}}$ are obtained from variable-temperature CW PL spectra (Figure S22). The data in Figure 4b show that $\tau_{^2\mathrm{E}}$ tracks $I_{^2\mathrm{E}}$ at low temperature, diverging somewhat above 28 K. Similarly, $\tau_{\mathrm{defectCr}}$ tracks $I_{\mathrm{defectCr}}$ closely, both dropping to nearly zero above 16 K. Notably, the $^2$E and defect $Cr^{3+}$ data do *not* track one another, with $\tau_{^2\mathrm{E}}$ and $I_{^2\mathrm{E}}$ changing more gradually over a broader temperature range than $\tau_{\mathrm{defectCr}}$ and $I_{\mathrm{defectCr}}$, indicating that these processes are not correlated in any straightforward way. These results confirm that this temperature dependence is dominated by changes in nonradiative processes.

We note that the same two PL signals and their different dependence on temperature were previously identified in CW PL measurements and assigned to the two low-symmetry-split spin-orbit components of the same $^2E$ state of the majority lattice $Cr^{3+}$ species.[34] Two observations argue against this interpretation: (*i*) the PLE data in Figure 2 show massively different excitation rates for the two emissive levels, with the lower-energy level not even detectable, whereas the two $^2E$ spinors of $Cr^{3+}$ should have similar absorption cross sections, and (*ii*) the time-resolved PL data in Figure 3 demonstrate that these two emissive states are not in thermal equilibrium with one another under any of our measurement conditions, as they should be for two $^2E$ sublevels of the same $Cr^{3+}$ ion. These considerations support assignment of the PL originating at ~10750 $cm^{-1}$ (~1.3328 eV) to a separate minority "defect" $Cr^{3+}$ site.

**Energy Migration and Exciton Dispersion.** Recently, we reported that doping $CrPS_4$ with $Yb^{3+}$ leads to efficient sensitization of $Yb^{3+}$ luminescence even at low $Yb^{3+}$ concentrations.[23] We now capitalize on this property by using $Yb^{3+}$ dopants to assess the dynamics of energy migration within the lattice. To illustrate, Figure 5a shows 36 K PL spectra of $CrPS_4$ crystals grown with and without $Yb^{3+}$ ($Yb_xCr_{1-x}PS_4$). With only $x = 0.002$, the $CrPS_4$ PL described above is almost completely replaced by narrow-line $Yb^{3+}$ *f-f* emission centered at ~9600 $cm^{-1}$ (~1.1903 eV), and the total PLQY ($\Phi'_{\mathrm{tot}}$) increases ~50×, now exceeding 25% above ~30 K (Figures S23 and S24). This large increase in PLQY with doping reinforces the conclusion drawn above that the lattice $Cr^{3+}$ PL in $CrPS_4$ is strongly suppressed by nonradiative processes. The $Yb^{3+}$ PL accounts for ~97.6% of all emission from this sample at 4 K, rising to >98.1% at 30 K and above. The shape of the lattice $^2E$ PL remains very similar to that of undoped $CrPS_4$ (Figure S22) regardless of *x*. These observations indicate rapid energy migration among lattice sites, analysis of which can provide information about the intrinsic luminescent excited state(s) of $CrPS_4$, which to date has been described as a localized $^2E$ state of a single $Cr^{3+}$ ion.

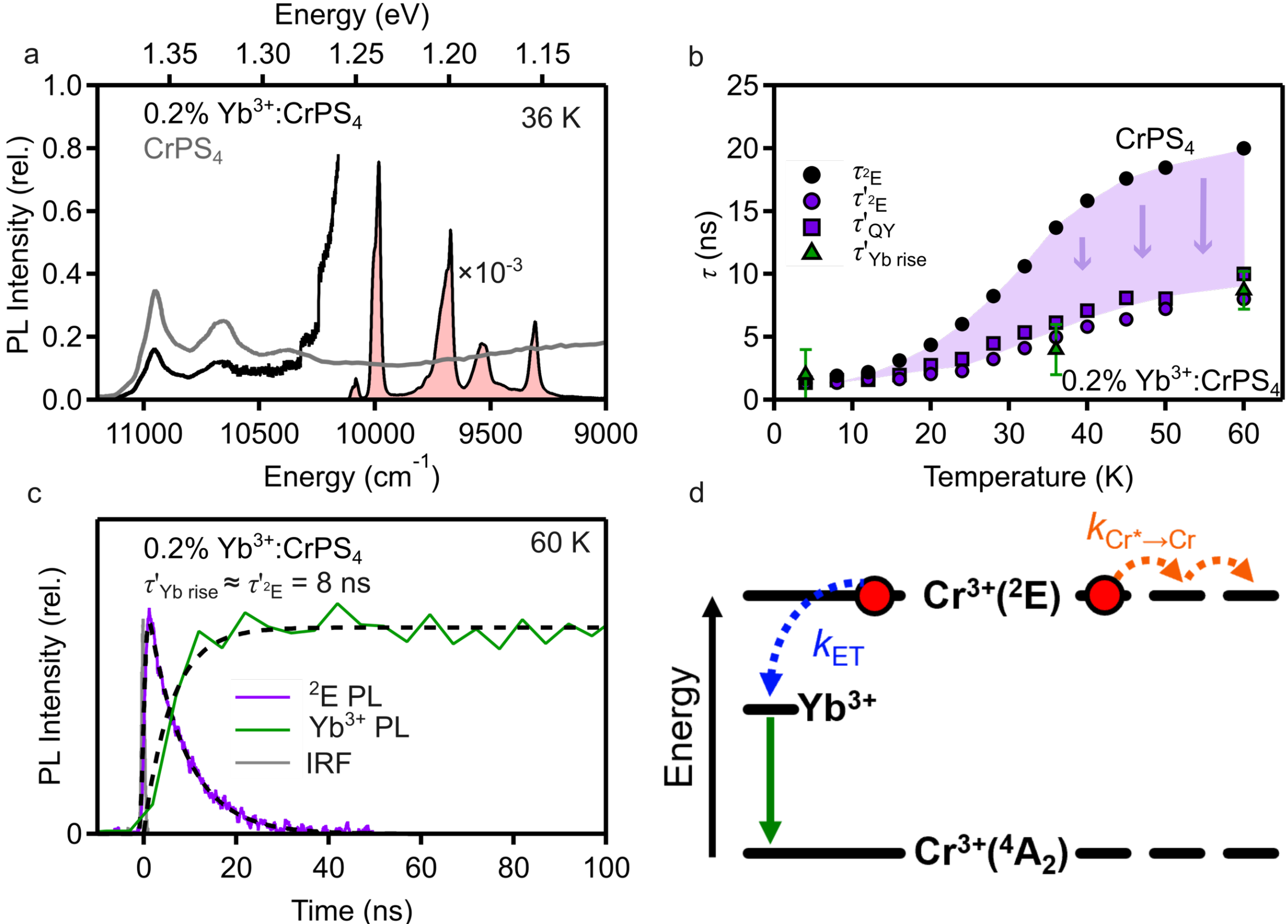

**Figure 5.** **(a)** 36 K PL spectrum of a single crystal of 0.2% $Yb^{3+}$:$CrPS_4$ (black) and $CrPS_4$ (gray, redrawn from Figure 2) relative to the PL intensities at 298 K (Figure S13). **(b)** Temperature dependence of $\tau_{^2E}$ (black circles, undoped, replotted from Figure 4b), $\tau'_{^2E}$ (purple circles, doped), $\tau'_{QY}$ (purple squares, doped) *cf.* eq 4, and $\tau'_{Yb\ rise}$ (green triangles with error bars). **(c)** 60 K time-resolved $Yb^{3+}$ PL curve of 0.2% $Yb^{3+}$:$CrPS_4$ measured at 10959 $cm^{-1}$ (spectral bandwidth (sbw) ±15 $cm^{-1}$) by omitting the deep-trap PL (green) collected by photoexciting with the pulsed 18797 $cm^{-1}$ (532 nm) output of a frequency-doubled Nd:YAG laser (unpolarized, 50 Hz, 30 ps pulse duration, 4 µJ pulse energy). The data are fitted (dashed black) to a monoexponential rise ($\tau'_{Yb\ rise}$ = 8.7 ns) and monoexponential decay ($\tau'_{Yb\ decay}$ = 83 µs). 60 K intrinsic lattice $^2$E PL decay curve of 0.2% $Yb^{3+}$:$CrPS_4$ measured at 9302 $cm^{-1}$ (sbw ±11 $cm^{-1}$, purple) collected and fitted (dashed black, $\tau'_{^2E}$ = 8.0 ns) as described in Figure 4a. **(d)** Illustration of the kinetic parameters discussed in the main text: $k_{ET}$ denotes the phenomenological $CrPS_4 \to Yb^{3+}$ energy-transfer rate constant, and $k_{Cr^* \to Cr}$ denotes the microscopic $Cr^{3+*} \to Cr^{3+}$ excitation hopping rate constant.

Time-domain PL measurements were used to learn more about energy migration in $CrPS_4$. Figure 5b compares $Cr^{3+}$ PL decay times ($\tau'_{^2E}$) measured for 0.2% $Yb^{3+}$:$CrPS_4$ with those for the same lattice $^2$E emission of undoped $CrPS_4$ ($\tau_{^2E}$ from Figure 4b). Both decay times increase with increasing temperature in a similar way, but PL decay is ~2× faster in the doped sample, indicating that energy transfer from $CrPS_4$ to $Yb^{3+}$ ($k_{ET}$) occurs on the same timescale as the PL decay (~2 ns at 4 K). Figure 5c plots time-resolved $Yb^{3+}$ PL measured at 60 K after pulsed excitation of the $CrPS_4$ lattice and omitting the trap PL (see Figures S26 and S27). The PL does not decay immediately but instead shows a distinct rise at short times. The fitted rise time of $\tau'_{\mathrm{Yb\,rise}}$ = 8.7 ns agrees well with the lattice $^2$E PL decay time measured under the same conditions ($\tau'_{^2\mathrm{E}}$ = 8.0 ns). $\tau'_{\mathrm{Yb\,rise}}$ values measured in this way at 4, 36, and 60 K are plotted in Figure 5b and track the $Cr^{3+}$ $^2$E decay data precisely. Very similar results are also obtained from analysis of the change in lattice $Cr^{3+}$ $^2$E PLQY upon $Yb^{3+}$ doping, as described by eq 4 and plotted as $\tau'_{\mathrm{QY}}$ in Figure 5b. From these data, we can directly attribute the decrease of the lattice $^2$E PL decay time and quantum yield upon $Yb^{3+}$ doping to capture of excitation energy by $Yb^{3+}$.

$$\tau'_{\mathrm{QY}} = \tau_{^2\mathrm{E}} \cdot \frac{\Phi'_{^2\mathrm{E}}}{\Phi_{^2\mathrm{E}}} \tag{4}$$

At such low $Yb^{3+}$ doping levels, the $CrPS_4$ excitation typically occurs far from any $Yb^{3+}$, and energy migration within the lattice is required before capture by $Yb^{3+}$. To analyze the microscopic $Cr^{3+}$–$Cr^{3+}$ hopping rate constant ($k_{\mathrm{Cr^*\rightarrow Cr}}$), we first determine the phenomenological $CrPS_4 \rightarrow Yb^{3+}$ energy-transfer rate constant ($k_{\mathrm{ET}}$) based on the Inokuti–Hirayama treatment of donor decay modified by energy transfer (eq 5).[57,58]

$$k_{\mathrm{ET}} = \frac{1}{\tau'_{^2\mathrm{E}}} - \frac{1}{\tau_{^2\mathrm{E}}} \tag{5}$$

$k_{\mathrm{ET}}$ decreases with increasing temperature from 0.15 ns$^{-1}$ at 4 K to 0.07 ns$^{-1}$ at 60 K (Figure S25a). Similar results are obtained when using PL quantum yields to calculate $k_{\mathrm{ET}}$ (eq 6, see Figure S25a).

$$k_{\mathrm{ET}} = \frac{1}{\tau_{^2\mathrm{E}}} \cdot \left(\frac{\Phi_{^2\mathrm{E}}}{\Phi'_{^2\mathrm{E}}} - 1\right) \tag{6}$$

For energy migration dominated by the strong intra-chain coupling between $Cr^{3+}$ ions at the smallest $Cr^{3+}$–$Cr^{3+}$ distances (relative to inter-chain or inter-layer coupling over larger distances), a microscopic rate constant for diffusive 1-D hopping ($k_{\mathrm{Cr^*\rightarrow Cr}}$) can be estimated from $k_{\mathrm{ET}}$ as described by eq 7.[59,60]

$$k_{\mathrm{ET}} = k_{\mathrm{Cr^*\to Cr}}[4x]^2 \quad (7)$$

From eq 7 and the 4 K value of $k_{\mathrm{ET}}$ = 0.15 ns$^{-1}$ for 0.2% $Yb^{3+}$:$CrPS_4$ ($x$ = 0.002), we obtain $k_{\mathrm{Cr^*\to Cr}}$ = 2.4 ps$^{-1}$. Similar values are obtained for all $Yb^{3+}$ concentrations examined here (Figure S25). We note that because $k_{\mathrm{Cr^*\to Cr}}$ is determined from energy capture at low $Yb^{3+}$ doping concentrations, where energy-migration lengths are large, it represents an effective value despite the presence of two crystallographic $Cr^{3+}$ sites. Figure 5d illustrates $k_{\mathrm{Cr^*\to Cr}}$ and $k_{\mathrm{ET}}$, summarizing the exciton dynamics analyzed here. Overall, these data indicate sub-picosecond $Cr^{3+}$–$Cr^{3+}$ hopping times.

This inter-site hopping in $CrPS_4$ is extremely fast compared with that determined for $CrI_3$ ($k_{\mathrm{Cr^*\to Cr}}$ ~ 10 μs$^{-1}$),[61] where hopping is slowed by strong electron-nuclear coupling in the emissive $^4T_2$ excited state. This hopping is too fast to be mediated by multipolar coupling,[62,63] and instead, it is attributed to inter-site exchange (Dexter-type energy transfer). Substantial $Cr^{3+}$–$Cr^{3+}$ exchange coupling is evident from the magnetic ordering of $CrPS_4$. Importantly, with this coupling, excited states consisting of just one $Cr^{3+}$ ion in its $^2E$ excited state are no longer eigenstates of the periodic lattice, and instead the eigenstates are Frenkel excitons comprising symmetrized linear combinations of single-ion $^2E$ excited states. This coupling gives rise to exciton dispersion and, in materials hosting translationally inequivalent neighboring ions, Davydov-type resonant energy splittings at the Brillouin-zone center (**k** = 0).[64-66] The inequivalent neighboring $Cr^{3+}$ ions in $CrPS_4$ show much stronger coupling than the equivalent next-nearest-neighbors, and $^2E$ energies are ligand-field independent, so the Davydov splittings ($\Delta E_{\mathrm{Dav}}$) should provide a good measure of the total exciton dispersion energy ($\Delta E_{\mathrm{band}}$).[66]

Experimentally, Davydov splittings are not resolved in either our PL or laser PLE measurements, but their magnitude can be estimated from consideration of the experimental energy migration rates measured above. In a symmetrical $Cr^{3+}$ Frenkel dimer, $\Delta E_{\mathrm{Dav}}$ is related to the coherent inter-site energy-transfer rate constant $k_{\mathrm{transfer}}$ as described by eq 8a,[47,67]

$$\Delta E_{\mathrm{Dav}} = \left(\frac{h}{2}\right) k_{transfer} \quad (8a)$$

where $h$ is Planck's constant. Extension to an infinite linear chain is described by eq 8b.[68]

$$\Delta E_{\mathrm{Dav}} = \left(\frac{h}{4}\right) k_{transfer} \tag{8b}$$

From eq 8b and assuming $k_{\mathrm{transfer}} = k_{\mathrm{Cr^*\rightarrow Cr}}$ = 2.4 ps$^{-1}$ at 4 K, a value of $\Delta E_{\mathrm{Dav}}$~ 20 cm$^{-1}$ (~2.5 meV) would be estimated for the $^2$E exciton of $CrPS_4$ at this temperature. $\Delta E_{\mathrm{Dav}}$ is thus small, and even this value is likely an overestimation. A more precise estimate of $\Delta E_{\mathrm{Dav}}$ would require accounting for the presence of two lattice sites, vibronic coupling, and any common structural disorder in the lattice. These factors all favor excitation localization and loss of coherence, opposing the dispersive electron-exchange interactions. Additionally, the roles of inter-chain or inter-layer hopping may be underestimated by assuming a linear-chain model.

**Exciton-Magnon Coupling and Other Spin Effects.** Mirrored fine structure between absorption and PL spectra of transition-metal ions is commonly observed due to electron-nuclear coupling, where vibronic transitions appear as sidebands to pure electronic transitions. Such a picture is described well by ligand-field theory and has already been discussed for $CrPS_4$,[28,29,34-36,39-44] but this picture is only adequate for isolated transition-metal ions. For $CrPS_4$, the optical spectra are also expected to show concrete manifestations of inter-ion magnetic exchange coupling. In particular, coupling of $^4A_2 \leftrightarrow {}^2E$ spin-flip transitions with dispersive magnons introduces a new mechanism for conserving spin angular momentum, and therefore generating electric-dipole allowedness in the optical transitions.[66,69-73] Tanabe and coworkers[74,75] have detailed an exchange mechanism by which spin-forbidden *d-d* transitions of exchange-coupled ions can gain electric-dipole allowedness. The dominant terms contributing to electric-dipole intensity in this mechanism involve the same inter-ion one-electron transfer processes as in the ground-state magnetic-exchange coupling. This "Tanabe mechanism" is responsible for intensification of $^4A_2 \leftrightarrow {}^2E$ transitions in the optical spectra of exchange-coupled $Cr^{3+}$ pairs in doped crystals and molecules, following the pair spin selection rules $\Delta S = 0$ and $\Delta M_S = 0$, where $S$ is the total pair spin ($S_{\mathrm{pair}}$), and it also accounts well for the properties of ions in fully concentrated magnetic crystals.[74-78] When exciton dispersion is small, as found here, the magnetically activated sidebands should closely resemble the magnon density of states (DOS), with intensities weighted by the momentum-resolved coupling strength.

To investigate whether any of the fine structure in the PL or PLE spectra of $CrPS_4$ derives from magnetic exchange, the magnon DOS was calculated (see Methods). The magnetic Hamiltonian used here is given by eq 9,[30]

$$\mathcal{H} = -2J_1 \sum_{i,j} \mathbf{S}_i \cdot \mathbf{S}_j - 2J_2 \sum_{i,k} \mathbf{S}_i \cdot \mathbf{S}_k - 2J_3 \sum_{i,l} \mathbf{S}_i \cdot \mathbf{S}_l - 2J_c \sum_{i,m} \mathbf{S}_i \cdot \mathbf{S}_m - \sum_{i,z} D_z (S_i^z)^2 \qquad (9)$$

where $J_n$ are the magnetic exchange coupling parameters ($J_1$ and $J_2$ describe the alternating nearest-neighbor pairwise interactions along the *b* axis, $J_3$ describes the nearest-neighbor interaction along the *a* axis, and $J_c$ describes the nearest-neighbor interaction along the *c* axis) and $D_z$ is the effective $Cr^{3+}$ single-ion anisotropy. The relevant $J_n$ values were obtained from experimental inelastic neutron scattering (INS) data[30] (accounting for the use of a $+J$ Hamiltonian in the source publication).

In the low-temperature limit, $^4A_2 \leftrightarrow {}^2E$ transitions can only couple with magnon excitation, not magnon de-excitation, because no thermally excited magnons exist. Consequently, the lowest-energy feature in the absorption/PLE spectrum will be the pure electronic transition (at $E(^2E)$) and any magnon sidebands must appear to higher energy. Similarly, the highest energy feature in the low-temperature PL spectrum will be the electronic origin (again at $E(^2E)$) and any magnon sidebands must appear to lower energy. For any magnon sideband that appears in both spectra, the average energy of PL and PLE sideband peaks is $1/2([E(^2E) + E(\text{magnon})] + [E(^2E) - E(\text{magnon})]) = E(^2E)$ (see Figure S30). From the energies of the mirrored magnon sideband features in the PL and PLE spectra, a value of $E(^2E) = 11069$ cm$^{-1}$ (1.3724 eV) was thus determined.

Figure 6a replots the low-temperature PL and PLE spectra from Figure 1d with the magnon DOS superimposed in both absorption and emission directions. The dashed vertical line indicates the common pure-electronic origin ($E(^2E)$). The magnon DOS maxima, calculated with no adjustable parameters, align remarkably well with the low-energy fine-structure peaks observed in both PL and PLE. This comparison provides strong evidence that the *majority* of this fine structure indeed comes from exciton-magnon coupling. We note that some of this fine structure in the PL spectrum was previously interpreted in terms of a Fano-type resonance involving coupling of single-ion $^2E$ PL to an atom-like defect transition,[29,34,43] but given the mirrored fine structure in the PLE spectrum, this interpretation would require a second Fano-type resonance with another atom-like defect transition at a precisely equivalent energy *above* the $Cr^{3+}$ electronic origin, which does not seem likely. We assert that the prediction of this fine structure from entirely independent INS data[30] provides a more compelling interpretation of this structure in terms of magnon sidebands to the $^4A_2 \leftrightarrow {}^2E$ spin-flip electronic transitions. Additionally, the good agreement between the optically detected magnon DOS and that measured by INS demonstrates that the magnon energies are very similar in the samples used in these two experiments. Importantly, all of the magnon DOS maxima in Figure 6a stem from dispersion primarily along the crystallographic *b* direction. For example, the pair of peaks ~80 cm$^{-1}$ from the origin is a signature of the magnon DOS gap that is formed by having two alternating *J* values along the *b*-oriented 1D chains of $Cr^{3+}$ ions ($J_1$ and $J_2$). Because the primary source of electric-dipole intensity in this transition is $Cr^{3+}$-$Cr^{3+}$ exchange, we can conclude that the magnetic anisotropy of $CrPS_4$ is ultimately responsible for the optical polarization along *b*.

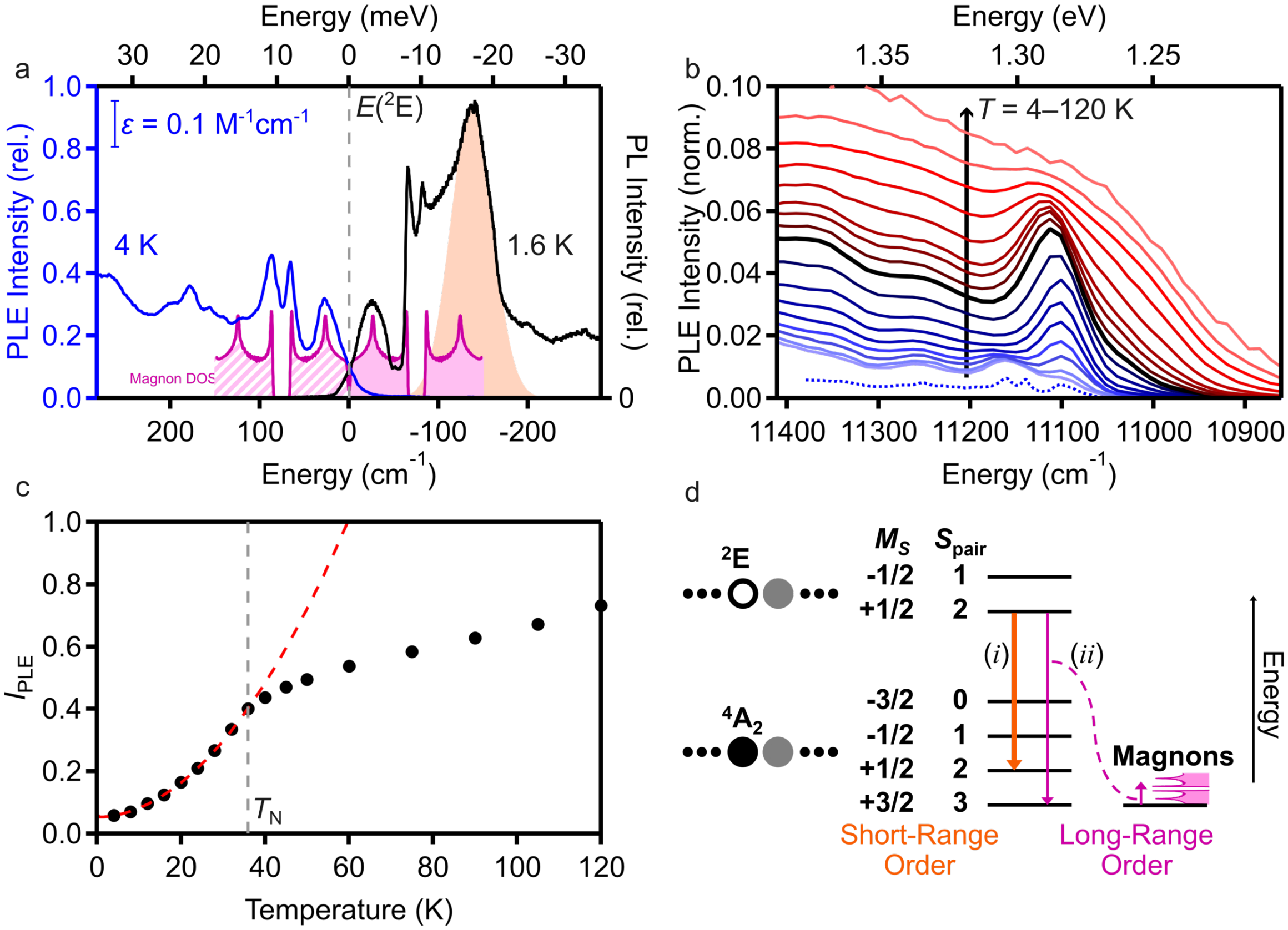

**Figure 6. (a)** 1.6 K PL (black) and 4 K PLE (blue) spectra redrawn from Figure 1d and offset to center $E(^2E)$ at zero (gray dashed line). The magnon DOS (magenta, arb. scaling) is plotted to positive (partial fill) and negative (solid fill) energies. The spin-preserving $^2E(+1/2, 2) \rightarrow {}^4A_2(+1/2, 2)$ transition is shaded orange. The scale bar indicates the molar extinction coefficient ($\varepsilon$), determined from absorption spectroscopy (see Figures 1c and S11). **(b)** Variable-temperature PLE spectra of exfoliated 0.2% $Yb^{3+}$:$CrPS_4$ crystals on clear adhesive tape (see Methods) in the region of the $^4A_2 \rightarrow {}^2E$ transition. Spectra are collected using a CW halogen lamp for photoexcitation (10-50 μW/cm$^2$, 1 nm linewidth) with $Yb^{3+}$ PL detection at 9302 cm$^{-1}$ (sbw ±22 cm$^{-1}$) and normalized to the maximum PLE intensity of the $^4A_2 \rightarrow {}^4T_2$ transition (Figure S28). The 36 K spectrum is emphasized (thick black). The high-resolution PLE spectrum from panel (a) is redrawn for comparison (dashed blue). **(c)** Integrated PLE intensity of the $^4A_2 \rightarrow {}^2E$ transition *vs* temperature from the data in panel (b). Data were integrated from 11000–11200 cm$^{-1}$ (1.3638–1.3886 eV). The dashed curve describes $T^2$ temperature dependence described by eq 10. **(d)** Schematic depiction of an exchange-coupled $Cr^{3+}$ pair within an extended $CrPS_4$ linear chain. Electronic excitation localized around one ion of the pair is depicted. Energy diagram describing the two exchange-mediated spin-conserving PL mechanisms identified in the $Cr^{3+}$ $^2E \rightarrow {}^4A_2$ fine structure of $CrPS_4$: (*i*) a $^2E(+1/2, 2) \rightarrow {}^4A_2(+1/2, 2)$ transition that preserves the total spin and spin projection *via* short-range magnetic order within pairs, and (*ii*) magnon-assisted transitions that couple the spin-forbidden $^2E(+1/2, 2) \rightarrow {}^4A_2(+3/2, 3)$ transition to compensating spin-wave excitations (long-range order). Energies not to scale.

An interesting aspect of these data is that magnon sidebands are observed at energies of both $E(^2E) + E(\text{magnon})$ (in PLE) and $E(^2E) - E(\text{magnon})$ (in PL) at the lowest measurement temperature. The observation by PLE of such magnon sidebands would be unexpected for monolayer $CrPS_4$, because long-range ferromagnetic spin correlation at the lowest temperature precludes magnon creation that *increases* spin angular momentum to offset the loss of spin angular momentum in the $^4A_2 \rightarrow {}^2E$ optical excitation. Such processes do become possible in layered $CrPS_4$, however, due to the antiferromagnetic *inter*-layer coupling. In contrast, the magnon sidebands in the low-temperature PL spectrum couple the gain in spin angular momentum from the $^2E \rightarrow {}^4A_2$ transition with loss of spin angular momentum from *intra*-layer magnon creation.

Additional evidence of exciton-magnon coupling is observed in variable-temperature PLE measurements. Figure 6b plots PLE spectra collected while monitoring the $Yb^{3+}$ *f-f* PL of a 0.2% $Yb^{3+}$:$CrPS_4$ flake. To eliminate any internal temperature dependence of the $Yb^{3+}$ PL, these PLE intensities are plotted relative to the $^4T_2$ peak maximum of each PLE spectrum (Figure S28), based on the observation in Figure 1c (and Figure S29) that $^4T_2$ absorption changes little over this temperature range (as expected for a spin-allowed *d-d* transition in low symmetry). The narrowest linewidths in these spectra are instrument-limited (see Figure 6a), but the trends are nonetheless apparent: as the temperature increases from 4 K, the $^2E$ PLE intensity increases dramatically and its structure broadens. Figure 6c plots the temperature dependence of this PLE intensity integrated from 11000–11200 cm$^{-1}$ (1.3638–1.3886 eV), showing that it increases rapidly above 4 K. This strong temperature dependence continues until $T_N \sim 36$ K and then abruptly changes to a weaker temperature dependence, demonstrating that it is linked to $CrPS_4$ magnetic order. Indeed, this temperature dependence is very similar to that observed in the magnetic exchange splittings of

$Yb^{3+}$ dopant PL lines in $CrPS_4$ (Figure S28d).[23]

Microscopically, this PLE temperature dependence from 4 K – $T_N$ reflects the introduction of new spin-conserving transitions that couple $Cr^{3+}$ $^4A_2 \rightarrow {}^2E$ excitation with hot-magnon deexcitation. The data are described well in terms of the loss of long-range spin correlation in $CrPS_4$ *via* thermal magnon excitation, following the expected $T^2$ dependence described by eq 10.[79]

$$I(T) = I_0 + AT^2 \exp\left(\frac{-\Delta_0}{k_{\mathrm{B}}T}\right) \tag{10}$$

Here, $I_0$ is the intensity at 0 K, $A$ is a scalar, $\Delta_0$ is the spin-wave gap (1.05 $cm^{-1}$ (0.13 meV) measured in ref 30), and $k_B$ is the Boltzmann constant. Because $\Delta_0 << k_B T$ in all measurements, this temperature dependence essentially follows $T^2$. This process is thus similar to the hot magnon sidebands of optical spin-flip transitions in easy-plane $A_2CrCl_4$ layered ferromagnets.[80,81] As described above, magnons coupled to $^4A_2 \rightarrow {}^2E$ excitations must all *increase* the lattice spin projection, and in the low-temperature limit of $CrPS_4$ this can only be achieved by exciting magnons across the van der Waals gap. Raising the temperature introduces the possibility of coupling with primarily *intra*-layer hot magnons, and the large PLE intensity increase in Figure 6b,c indicates that this is a more effective intensity-gaining mechanism than inter-layer magnon coupling, consistent with the dependence of electric-dipole intensity on exchange-coupling strength outlined in the Tanabe mechanism.

Finally, although several of the same fine-structure peaks appear in both the absorption (PLE) and PL spectra of $CrPS_4$, the PL spectrum shows an additional prominent peak at $E(^2E)$ - 140 $cm^{-1}$ that is absent from the PLE spectrum. This peak is also assigned as a transition that has become spin-allowed ($\Delta S = 0$) through exchange *via* the Tanabe mechanism, but in this case reflecting only short-range order. To illustrate, Figure 6d depicts the energy levels of an exchange-coupled $Cr^{3+}$ ion relaxing from its $^2E$ ($S$ = 1/2) excited state to its $^4A_2$ ($S$ = 3/2) ground state. Fixing all neighboring intralayer $Cr^{3+}$ spin projections to +3/2 in the ordered state, ferromagnetic exchange coupling between this central ion and the surrounding $Cr^{3+}$ ions yields the ground- and excited-state spin splittings depicted. The corresponding pair spin states ($S_{pair}$) involving the nearest-neighbor $Cr^{3+}$ are also indicated. Whereas the highest-energy $^2E \rightarrow {}^4A_2$ PL transition at low temperature (denoted $^2E(+1/2, 2) \rightarrow {}^4A_2(+3/2, 3)$, with $\Delta S_{pair} = +1$) requires a spin flip and must therefore couple with other lattice spins to gain electric-dipole allowedness as described above, the next lower-energy transition ($^2$E(+1/2, 2) → $^4$A$_2$(+1/2, 2), $\Delta S_{pair}$ = 0) has no such restriction. The far greater intensity of this $^2$E(+1/2, 2) → $^4$A$_2$(+1/2, 2) transition compared to any of the $^2$E(+1/2, 2) → $^4$A$_2$(+3/2, 3) magnon sidebands demonstrates the greater efficacy of short-range order for overcoming $^2$E → $^4$A$_2$ spin-forbiddenness. The PL peak at $E(^2\text{E})$ - 140 cm$^{-1}$ is thus assigned to transition (*i*) in Figure 6d, and its shift from the first electronic origin is related to the $^4$A$_2$(+1/2, 2) – $^4$A$_2$(+3/2, 3) ground-state exchange splitting. If the excited-state exchange coupling were instead antiferromagnetic, then this PL energy shift would reflect the splitting between $^4$A$_2$(+3/2, 3) and $^4$A$_2$(-1/2, 1) levels in Figure 6d. Notably, regardless of the sign of the excited-state exchange coupling, there is no absorption transition originating from the $^4$A$_2$(+3/2, 3) low-temperature ground state that conserves spin, and hence no comparably intense magnetic sideband in the PLE spectrum of Figure 6a. The $^4$A$_2$(+1/2, 2) → $^2$E(+1/2, 2) excitation should appear as a hot band, and sizeable hot-band intensity redshifted from the low-temperature origin is observed (Figure 6b), but with an energy gap of 140 cm$^{-1}$ and $T_N$ of only 36 K this transition would carry significant intensity only in the paramagnetic regime, where extensive broadening and other hot bands also occur. This interpretation of the $E(^2\text{E})$ - 140 cm$^{-1}$ PL feature is supported by the PL temperature dependence across the magnetic ordering temperature: at $T_N$ and above (*e.g.*, at 36 K, Figure 2), where long-range spin correlation vanishes but short-range exchange splittings remain, the magnon fine structure has broadened and disappeared whereas the $^2$E(+1/2, 2) → $^4$A$_2$(+1/2, 2) transition depicted in Figure 6d persists. This transition dominates the $^2$E PL intensity at these temperatures, serving as the magnetic "false" origin for the most prominent vibronic progression.

## III. Conclusion

The low-temperature PL spectrum of the layered van der Waals magnet $CrPS_4$ has long been known to display extremely complex fine structure. Although to date interpreted in terms of the single-ion ligand-field model as $^2$E → $^4$A$_2$ luminescence, the magnetic ordering of this lattice implies a role for magnetism in its PL that has not been previously accounted for. Here, we have presented PL measurements at various temperatures and in the time domain, in conjunction with absorption, PLE, and PL quantum yield measurements, that lead to several new conclusions about the optical spectroscopy and electronic structure of $CrPS_4$: (*a*) the intrinsic pure electronic transition occurs at 11069 cm$^{-1}$ (1.3724 eV), as determined from mirrored PL and PLE fine structure at low temperature, (*b*) the majority of the low-energy fine structure in both PL and PLE spectra can be assigned as magnon sidebands to the formally spin-forbidden $^4A_2 \leftrightarrow {}^2E$ electronic transitions of lattice $Cr^{3+}$ ions, gaining electric-dipole intensity through the Tanabe exchange mechanism, (*c*) the intrinsic PL intensity is dominated by a magnetic "false" origin 140 $cm^{-1}$ (17.4 meV) below the pure electronic transition and its associated vibronic sidebands, particularly at and above $T_N$, whose intensity also derives from the Tanabe mechanism but involves only short-range magnetic order, (*d*) PL from an additional "defect" $Cr^{3+}$ species is identified 208 $cm^{-1}$ (25.8 meV) below the lattice $Cr^{3+}$ PL using a combination of PLE and time-resolved PL measurements, and (*e*) sub-picosecond inter-site excitation hopping in $CrPS_4$ leads to Frenkel-type exciton formation with a small dispersion energy of $\Delta E_{band} < \sim 20$ $cm^{-1}$ (~2.5 meV). Several of these findings are summarized in Scheme 1.

**Scheme 1. Summary of the various dynamical processes occurring in $CrPS_4$ and $Yb^{3+}$:$CrPS_4$ following photoexcitation. Straight solid arrows indicate luminescence and dashed arrows indicate nonradiative processes. The five major radiative relaxation processes are (*i*) $^2E$ PL coupled to magnon excitation (magenta), (*ii*) $^2E$ PL terminating in an excited spin level of the exchange-split ground state (violet), (*iii*) $^2E$ PL from shallow "defect" $Cr^{3+}$ (blue), (*iv*) native deep-trap PL (green), and (*v*) $Yb^{3+}$ *f-f* PL in samples where $Yb^{3+}$ has been doped into $CrPS_4$ (orange). (*vi*) shows picosecond hopping between translationally inequivalent $Cr^{3+}$ ions.**

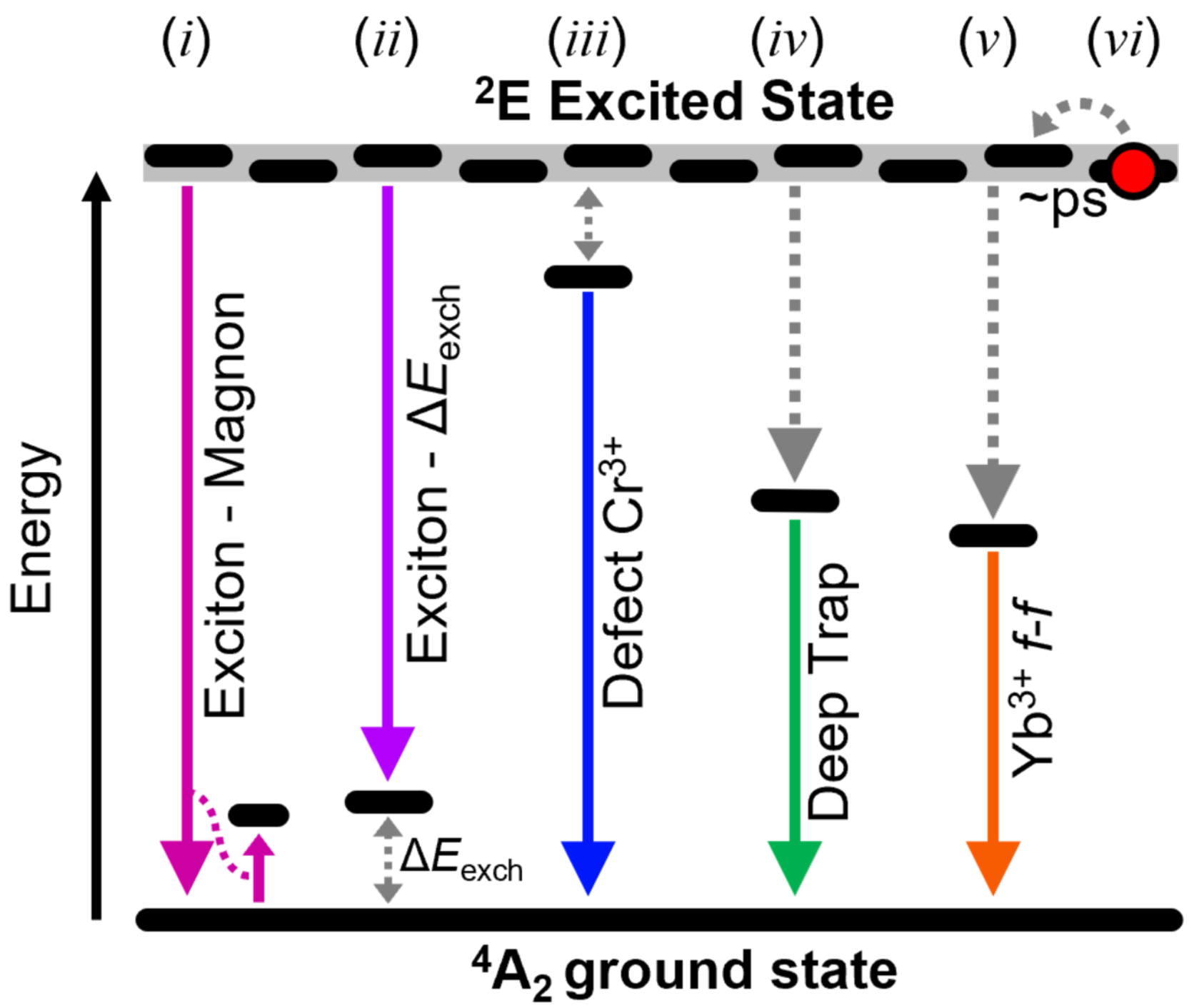

These findings demonstrate that most of the extremely rich and unusual structure in the optical spectra of $CrPS_4$, as well as its *b*-axis polarization, derives directly from this lattice's magnetic structure. Whereas magnon sidebands and exchange-enhanced spin-flip transitions have been detailed previously in 3D transition-metal ionic and oxide antiferromagnets,[56,65,66,69-73,76-78,82] such phenomena have not been widely observed in the optical spectroscopy of van der Waals magnets. To our knowledge, the only direct observation of magnon sidebands in the optical spectrum of a 2D thiophosphate magnet comes from second-harmonic generation measurements on $MnPS_3$ in the region of the sharp $^6A_1 \rightarrow {}^4A_1/{}^4E(G)$ ligand-field excitations.[83,84] Magnon sidebands are also observed in spin-flip excitations of layered 2D metal-halide perovskites of $Cr^{2+}$ and $Mn^{2+}$.[80,81,85] At the same time, magnon generation *via* optical excitation has been demonstrated in numerous related 2D materials where magnon sidebands have not been identified or assigned, for example in the A-type antiferromagnet CrSBr.[7,86]

As another $Cr^{3+}$-based quasi-1D A-AFM material, CrSBr is a particularly significant comparison for this work. This material's absorption/PLE and PL have been widely described[87,88] as involving band-to-band (single-particle) electronic transitions, but in fact the spectra are remarkably similar to those of $CrPS_4$. Although never described in this way before, the data and analysis presented here for $CrPS_4$ suggest that the emissive excited state of CrSBr could instead be a dispersive $^2E$ ligand-field state. Three important differences between CrSBr and $CrPS_4$ are: (*i*) the stronger $Cr^{3+}$–$Cr^{3+}$ exchange interactions, (*ii*) the presence of only one $Cr^{3+}$ lattice site, and (*iii*) the shorter (1- *vs* 3-atom) separation between 1D chains in CrSBr. If also $^2E$ emission, these differences would cause greater $^2E$ exciton dispersion both along and perpendicular to the primary 1D $Cr^{3+}$ chains, and hence possibly sizable deviation of the magnon sideband structure from the relatively unperturbed DOS energy pattern seen for $CrPS_4$. This comparison will be discussed further in a forthcoming report.

In conclusion, the spectroscopic results and analysis presented here shed new light on the role of magnetism in the optical transitions of $CrPS_4$, showing that the primary PL and absorption/PLE intensity-gaining processes all involve exchange-mediated coupling of $^4A_2 \leftrightarrow {}^2E$ ligand-field transitions with spin-compensating changes in the ground state, involving either magnons (long-range order) or local exchange splittings (short-range order) to overcome the strict spin selection rule ($\Delta S = 0$) of the single-ion spin-flip optical transition. This study advances our understanding of the fundamental electronic structure of $CrPS_4$ in ways that may help inform future studies and applications of magnetic van der Waals materials as a novel class of spin-photonic materials. Furthermore, the observation of resolved magnon sidebands in the PLE spectra of $CrPS_4$ points to the possibility of launching magnons optically with mode selectivity, offering new avenues for optically initiated spin generation and manipulation in layered magnetic materials.

## IV. Methods

**Synthesis of $CrPS_4$ and $Yb^{3+}$-doped $CrPS_4$ crystals.** Crystals were grown by chemical vapor transport in a manner similar to that described previously.[23,40] A chromium chip (99.995%, lot MKCH4484) was purchased from Sigma Aldrich. The Cr chip was ground to a powder using a mortar and pestle. Red phosphorus powder (> 97.0%, lot MKCS0319) was purchased from Sigma Aldrich. Sulfur powder (99.98%, lot STBL3899) was purchased from Sigma Aldrich. Ytterbium metal powder 40 mesh (99.9%) was purchased from BeanTown Chemical. All chemicals were used as received without further purification. In a typical synthesis, Yb metal ($x$ mmol), Cr metal (2.6 - $x$ mmol), P powder (2.6 mmol in P), and $S_8$ powder (1.3 mmol in S) were loaded (an $x$:1-$x$:1:4 mole ratio of the elements) into a quartz tube and sealed under an evacuated atmosphere. The quartz tubes were 10 cm long with inner and outer diameters of 14 and 16 mm, respectively. Sealed tubes were placed in an open-ended horizontal tube furnace with the starting materials in the hot zone set at 750°C and the other end at a temperature of 650°C. Samples were heated for 5 days and then allowed to cool to room temperature over a period of ca. 6 hr. Once cooled, the tubes were cracked open to yield shiny dark plate-like crystals that had formed at the cold end. $Yb^{3+}$ doping was confirmed by inductively coupled plasma mass spectrometry (ICP-MS) using a PerkinElmer NexION 2000B. ICP-MS samples were digested in high-purity concentrated nitric acid, followed by dilution in ultrapure $H_2O$. $Yb^{3+}$ doping levels ($x$) are reported as a percentage relative to the total cation content, $[Yb^{3+}]/([Cr^{3+}]+[Yb^{3+}])$. Analytical $Yb^{3+}$ concentrations were ~50 times smaller than the nominal ytterbium concentrations used in the synthesis. The measured $Yb^{3+}$ concentrations in the flakes studied here were 0.20 ± 0.05% and 0.020 ± 0.005% with the uncertainty accounting for flake-to-flake variability. The 0.002% $Yb^{3+}$ concentration is estimated based on the expected linear relationship between nominal concentration and analytical concentration because it was too dilute to measure. Crystal thickness was measured by mounting a flake to a glass slide using double-sided tape and imaging the edge of the flake with an optical microscope at various magnifications. The edge thickness was calculated in ImageJ2 using known pixel resolutions at several spots across the sample to obtain an average and range of thickness. $CrPS_4$ and 0.2% $Yb^{3+}$:$CrPS_4$ microcrystal solutions were prepared by 8-hr sonication of the parent single crystals in hexane.

**Magnetic measurements.** Magnetic measurements were collected on an as-synthesized single crystal using a Quantum Design MPMS®3 - SQUID Magnetometer (MEM-C Shared Facilities) collecting in the vibrating sample magnetometer mode.

**X-ray diffraction.** As-synthesized flakes were characterized by X-ray diffraction using a Bruker D8 Discover powder diffractometer with IμS microfocus X-ray source for Cu K$\alpha$ radiation (50 kV, 1 mA). Samples were placed onto crystalline silicon substrates and measured under ambient conditions.

**Raman scattering.** Raman measurements were performed using a Renishaw inVia Raman microscope equipped with a 1200 grooves/mm grating and a Leica DMIRBE inverted optical microscope. A 532 nm laser was used as the excitation source. The collimated monochromatic beam was focused onto the sample through a Leica objective lens. Raman-scattered light was collected by the same objective, directed into the inVia spectrometer, dispersed by the diffraction grating, and focused on a CCD camera.

**Transmission electron microscopy.** 8 μL of 0.2% $Yb^{3+}$:$CrPS_4$ microcrystal solution in hexane was drop-cast onto an Ultrathin Carbon Type A, 400 mesh Cu TEM grid three times, allowing the solvent to dry between drops. The grid was then dried overnight under vacuum before imaging with a Technai G2 F20 Supertwin TEM operating at 200 kV.

**Photoluminescence (PL).** Single crystals of the material were placed between two sapphire disks and loaded into a closed-cycle or helium immersion cryostat. Crystals were exfoliated prior to loading to ensure fresh surfaces. Measured samples were typically 100-500 μm thick. Measurements were performed under high vacuum ($10^{-6}$ Pa). Sample emission was focused into a monochromator with a spectral bandwidth of 0.05 nm (~0.5 $cm^{-1}$) for most measurements, and as specified in each figure. The excitation source and collection axes were both normal to the crystal face. Spectra were collected using a $LN_2$-cooled silicon CCD camera (measurements with < 1 nm spectral bandwidth) or a Hamamatsu InGaAs/InP NIR photomultiplier tube (measurements with ≥ 1 nm spectral bandwidth). All PL spectra were corrected for instrument response. Photoexcitation was performed using either a 632.8 nm helium-neon laser, a tunable Ti:sapphire laser, a 532 nm Nd:YAG laser, a 470 nm pulsed laser diode, a 405 nm LED, or a 375 nm pulsed laser diode as specified in the text. Where specified, polarization scrambling was achieved by passing the excitation source and sample emission through fused silica depolarizers. Unless specifically indicated otherwise, crystal orientation was not controlled relative to the excitation polarization, and the excitation source was depolarized.

**Photoluminescence excitation (PLE).** PLE measurements were likewise conducted in a closed-cycle helium cryostat under high vacuum. Sample emission was focused into a monochromator with a spectral bandwidth of 5 nm (~44 $cm^{-1}$) for $Yb^{3+}$:$CrPS_4$ or 15 nm (~150 $cm^{-1}$) for $CrPS_4$ and emission counts were monitored using a Hamamatsu InGaAs/InP NIR photomultiplier tube. Spectra were corrected for intensity of the excitation source, measured simultaneously using a calibrated Si diode. Photoexcitation was performed using either a halogen lamp coupled to a spectrometer with a 1 nm (~10 $cm^{-1}$) linewidth or a tunable Ti:sapphire laser with an integration-time-limited linewidth of 0.01 nm (~0.1 $cm^{-1}$). For Figures 5, S10, S11, and S28, "exfoliated 0.2% $Yb^{3+}$:$CrPS_4$ crystals on clear adhesive tape" were prepared by repeatedly exfoliating a 0.2% $Yb^{3+}$:$CrPS_4$ single crystal on the same piece of clear Scotch-brand tape until the tape was uniformly faintly orange. This piece of tape was mounted between two sapphire discs and measured as described above.

**Photoluminescence quantum yield (PLQY).** The PLQY was measured at room temperature for 11 different flakes, and the average was found to be 0.14 ± 0.04%, accounting for flake-to-flake variation and measurement variability. Large flakes (> 1 mm thick and 5 mm wide) must be used for PLQY due to the very weak luminescence, but small flakes (<0.5 mm thick and 2 mm wide) are used for low-temperature measurements to ensure effective cooling. The room-temperature PLQY is used to determine the PLQY at low temperature using the full vis-NIR spectra measured at low temperature and room temperature using a NIR PMT (Figure S13). Then, the high-resolution CCD spectra, despite missing the IR PL component, are analyzed by comparison with the low temperature PLQY in the overlapping measurement region (~9000–12000 cm$^{-1}$). PLQY for just the $^2$E luminescence ($\Phi_{^2\mathrm{E}}$) was determined by integrating the PL intensity from 10950–11200 cm$^{-1}$ and multiplying this intensity by 8 to represent the $^2$E PL (this integrated region is ~7–9× smaller than the entire deconvoluted $^2$E PL spectrum at 1.6 and 36 K).

**Spin-wave calculations.** Linear spin-wave calculations were performed using SpinW.[89] The crystallographic and magnetic parameters were obtained from ref 30. The $Cr^{3+}$ spins are assigned a 9.6° off-*c*-axis orientation based on ref 31. The magnon density of states was generated using 56-point mesh along each axis, 800 energy bins, and a 0.1 meV (0.8 cm$^{-1}$) Gaussian broadening factor. All four magnetic sublattices are included.

**Supplementary Materials.** See Supplemental Material at *[URL inserted by publisher]*. 30 figures showing supplemental XRD, Raman, and magnetic susceptibility data, absorption and PL data, optical microscope and TEM images, and brief descriptions of PLQY determination and energy-hopping dynamics.

**Data Availability.** The original experimental data presented in this work will be made freely available at https://doi.org/10.5281/zenodo.17643780 upon manuscript acceptance.

**Acknowledgments.** This research was primarily supported by the University of Washington Molecular Engineering Materials Center, a U.S. National Science Foundation Materials Research Science and Engineering Center (DMR-2308979). The work on $CrPS_4$ microcrystal solutions was supported by the U.S. National Science Foundation, Division of Materials Research (solid state and materials chemistry), through award DMR-2404703. Support from an NSF Graduate Research Fellowship (1000356387) to R.T.S. is gratefully acknowledged. Part of this work was conducted at the Molecular Analysis Facility, which is supported in part by funds from the Molecular Engineering & Sciences Institute, the Clean Energy Institute, the National Science Foundation (NNCI-2025489 and NNCI-1542101). The authors thank Dr. Kimo Pressler for assistance with the optical microscope measurements, Dr. Faris Horani for Raman measurements, Thom Snoeren for ICP-MS measurements, and Nick Adams for assistance with PL measurements.

## TOC Graphic

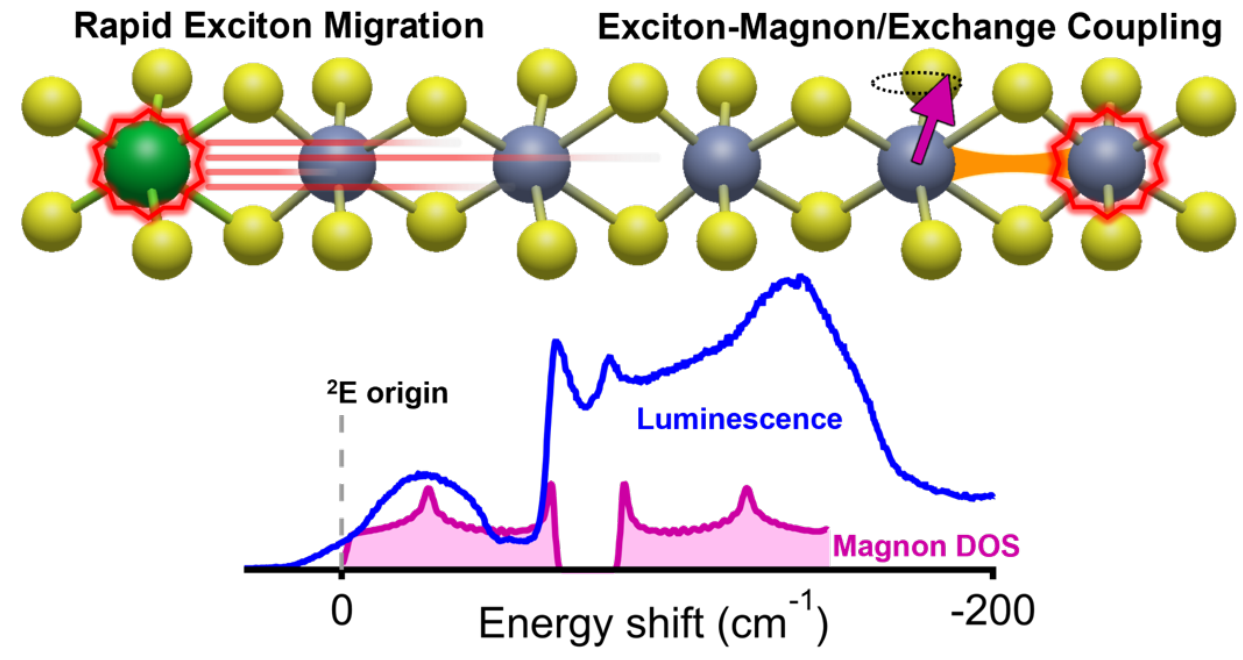